\documentclass[
prc,%
10pt,%
final,%
notitlepage,%
oneside,%
twocolumn,%
nobibnotes,%
nofootinbib,
superscriptaddress,%
floatfix,%
floatfix,%
showkeys,%
showpacs]%
{revtex4}
\usepackage{color} 
\usepackage{color}
\usepackage{amsfonts}
\usepackage{amsbsy}
\usepackage{mathrsfs}
\usepackage{graphicx}
\def\lsim{\mathrel{\rlap{
\lower4pt\hbox{\hskip-3pt$\sim$}}
    \raise1pt\hbox{$<$}}}     
\def\gsim{\mathrel{\rlap{
\lower4pt\hbox{\hskip-3pt$\sim$}}
    \raise1pt\hbox{$>$}}}     

\begin{document}
\title{	
Global $\Lambda$ polarization in moderately relativistic nuclear collisions
} 
\author{Yu. B. Ivanov}\thanks{e-mail: yivanov@theor.jinr.ru}
\affiliation{Bogoliubov Laboratory for Theoretical Physics, 
Joint Institute for Nuclear Research, Dubna 141980, Russia}
\affiliation{National Research Nuclear University "MEPhI", Moscow 115409, Russia}
\affiliation{National Research Center "Kurchatov Institute", Moscow 123182, Russia} 
\begin{abstract}
Predictions for the global polarization of $\Lambda$ hyperons in Au+Au collisions 
at moderately relativistic collision energies, 2.4 $\leq\sqrt{s_{NN}}\leq$ 11 GeV, are made. 
These are based on the thermodynamic approach to the global polarization
incorporated into the model of the three-fluid dynamics. 
Centrality dependence of the polarization is studied. 
It is predicted that the polarization reaches a maximum 
or a plateau (depending on the equation of state and centrality) at $\sqrt{s_{NN}}\approx$ 3 GeV. 
It is found that the global polarization increases with increasing width of the rapidity window around the midrapidity.  
\end{abstract}
\maketitle

Experimental discovery of global polarization of $\Lambda$ and $\bar{\Lambda}$  
hyperons in the STAR experiment \cite{STAR:2017ckg,Adam:2018ivw} 
gave us evidence of existence of a new class of 
collective phenomena in heavy-ion collisions.  
The global polarization of hyperons 
is well described within the thermodynamic approach based on hadronic degrees 
of freedom \cite{Becattini:2013fla,Becattini:2016gvu,Fang:2016vpj}.
This was demonstrated by implementation of this approach into 
various hydrodynamical 
\cite{Karpenko:2016jyx,Xie:2016fjj,Xie:2017upb,Ivanov:2019ern,Ivanov:2019wzg,Ivanov:2020wak}
and transport  
\cite{Li:2017slc,Sun:2017xhx,Wei:2018zfb,Shi:2017wpk,Kolomeitsev:2018svb,Vitiuk:2019rfv} 
models of heavy-ion collisions. 
This thermodynamic approach is not without its problems,   
see recent review in Ref. \cite{Becattini:2020ngo}. 
One of them, concerning the the global polarization, is that 
it does not explain large difference 
between $\Lambda$ and $\bar{\Lambda}$ polarization at 7.7 GeV. 
Various explanations of this $\Lambda$--$\bar{\Lambda}$ splitting
were put forward \cite{Vitiuk:2019rfv,Csernai:2018yok,Xie:2021fjn,Ayala:2020soy,Baznat:2017jfj,Ivanov:2020qqe}.
However, it is not yet quite clear how severe this problem of splitting is  
because the $\bar{\Lambda}$ polarization at 7.7 GeV is measured with poor accuracy.

At present, experiments at lower collision energies than those of the Beam Energy Scan (BES) program 
at the Relativistic Heavy Ion Collider (RHIC) 
are in progress:  STAR fixed target (FXT) program \cite{Meehan:2017cum} at RHIC 
and HADES \cite{Agakishiev:2009am} at GSI Helmholtzzentrum für Schwerionenforschung.
Also new facilities for heavy-ion collisions are under construction, 
which are designed for collider regime,  Nuclotron-based Ion Collider fAcility 
(NICA) in Dubna  \cite{Kekelidze:2017ghu}, and for fixed-target experiments,
the Baryonic Matter at Nuclotron (BM@N)\footnote{
In fact, the BM@N is already in operation, but the planed beams of really heavy ions with high luminosity 
will be achieved only in future.
} 
in Dubna \cite{Kapishin:2019wxa,BMN_CDR}, 
 the Facility for Antiproton and Ion Research (FAIR) in Darmstadt \cite{Ablyazimov:2017guv}, and
High Intensity heavy-ion Accelerator Facility
(HIAF) in Huizhou, China \cite{Lu:2016htm}.
The energy ranges for the Au beams are 
$\sqrt{s_{NN}}$ = 3 -- 7.2 GeV for STAR-FXT, 
$\sqrt{s_{NN}}$ = 2.3 -- 2.6 GeV for HADES, 
$\sqrt{s_{NN}}$ = 2.3 -- 3.5 GeV for BM@N, 
$\sqrt{s_{NN}}$ = 2.3 -- 4 GeV for HIAF,
$\sqrt{s_{NN}}$ = 2.7 -- 4.9 GeV for FAIR, and  
$\sqrt{s_{NN}}$ = 4 -- 11 GeV for NICA.  
Measurements of the global $\Lambda$ polarization are already in progress at the  
STAR-FXT and HADES \cite{Kornas:2020qzi}.

The NICA and BES-RHIC energy ranges overlap. Therefore, all above mentioned calculations of the 
global polarization 
\cite{Karpenko:2016jyx,Xie:2016fjj,Xie:2017upb,Ivanov:2019ern,Ivanov:2019wzg,Ivanov:2020wak,
Li:2017slc,Sun:2017xhx,Wei:2018zfb,Shi:2017wpk,Kolomeitsev:2018svb,Vitiuk:2019rfv,Baznat:2017jfj,Ivanov:2020qqe} 
involved the high-energy end of the NICA range. Even some works dedicated to the polarization at NICA and FAIR 
energies \cite{Xie:2016fjj,Kolomeitsev:2018svb} in fact considered  the upper end of this range that 
overlaps with BES-RHIC energies. 
Predictions of the global polarization in the actual NICA range 
were done in Refs. \cite{Ivanov:2019ern,Ivanov:2019wzg,Ivanov:2020wak,Baznat:2017jfj,Ivanov:2020qqe}. 
The first estimates at even lower energies but in terms of vorticity 
were recently reported in Ref. \cite{Deng:2020ygd}. All the estimates agree with that the 
global polarization (vorticity \cite{Deng:2020ygd}) rises with the collision energy 
decrease. Ref. \cite{Deng:2020ygd} predicts that the maximum vorticity is reached at 
$\sqrt{s_{NN}}\approx$ 3 GeV and then it decreases with decreasing collision energy.

In this paper, predictions for the global polarization of $\Lambda$ 
hyperons at HADES--NICA energies are made, based on the thermodynamic approach 
\cite{Becattini:2013fla,Becattini:2016gvu,Fang:2016vpj}.  
Simulations are performed within the model of the three-fluid dynamics (3FD) \cite{3FD}. 
The  3FD model takes into account nonequilibrium at the early stage of nuclear collisions.  
This nonequilibrium stage 
is modeled by means of two counterstreaming baryon-rich fluids. 
Newly produced particles, dominantly populating the midrapidity region, 
are attributed to a fireball fluid.
These fluids are governed by conventional hydrodynamic equations 
coupled by friction terms in the right-hand sides of the Euler equations. 
Calculations are done with three different equations of state  (EoS's): 
a purely hadronic EoS \cite{gasEOS}  
and two versions of the EoS with the   deconfinement
 transition \cite{Toneev06}, i.e. a first-order phase transition (1PT)  
and a crossover one. 
The physical input of the present 3FD calculations is described in
Ref.~\cite{Ivanov:2013wha}.

\begin{figure}[bht]
\includegraphics[width=8.1cm]{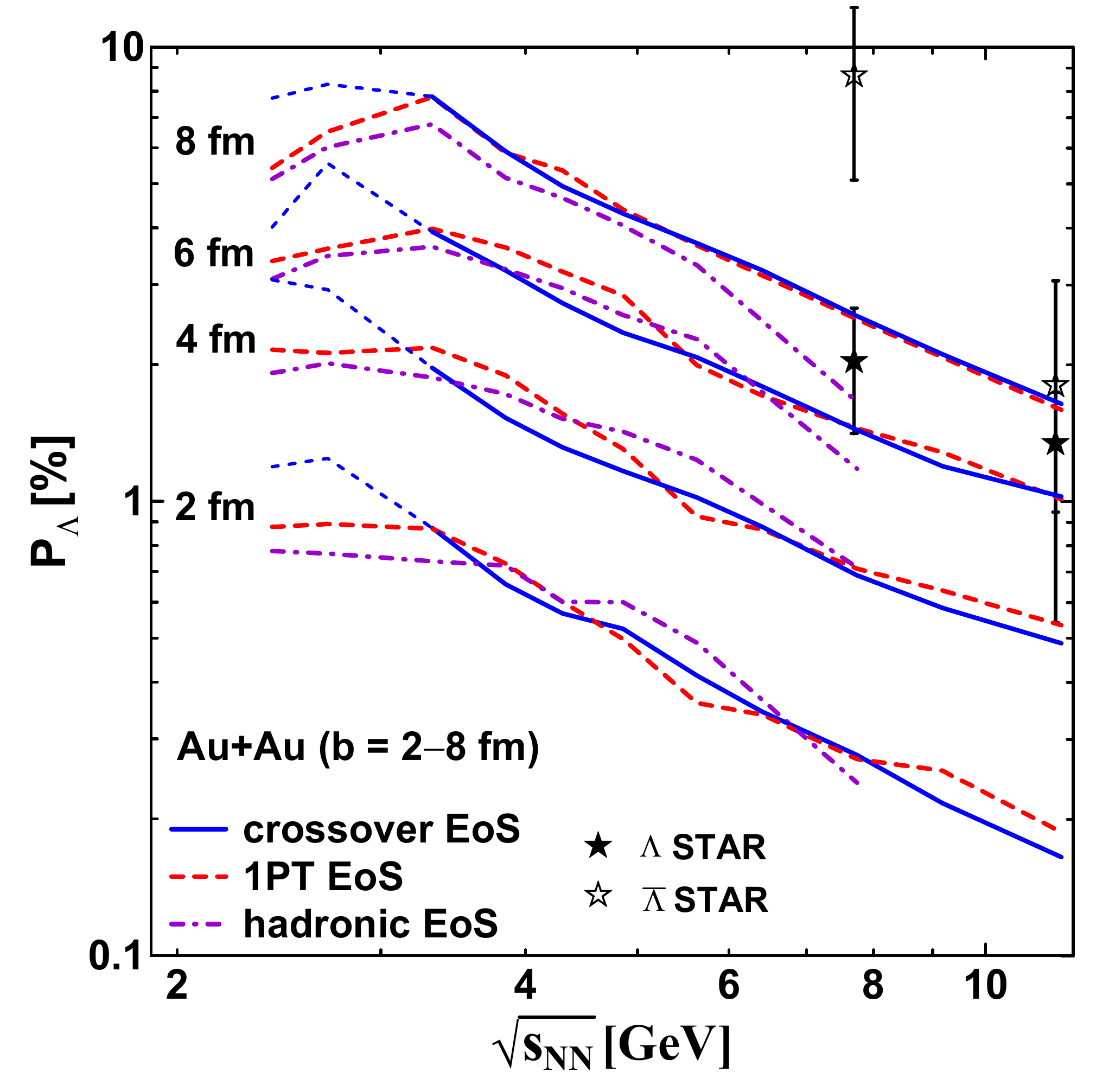}
 \caption{(Color online)
 Global polarization of $\Lambda$ hyperons in Au+Au collisions at impact parameters $b=$ 2, 4, 6, 8 fm as 
function of collision energy $\sqrt{s_{NN}}$ calculated with different EoS's 
within the framework of the thermodynamical approach.    
Short-dashed lines for the crossover EoS below 3 GeV indicate the unstable numerics. 
STAR data on global $\Lambda$ and $\bar{\Lambda}$ polarization \cite{STAR:2017ckg} 
are also displayed. 
}
\label{fig1}
\end{figure}

In Fig. \ref{fig1} the global polarization of $\Lambda$ hyperons in Au+Au 
collisions at different centralities, i.e. impact parameters $b=$ 2, 4, 6 and 8 fm, 
is presented in the collision-energy range extended to the low energies. 
The STAR data \cite{STAR:2017ckg} are also displayed to 
connect the low-energy predictions  with the data available at BES-RHIC. 
The impact parameter $b=$ 8 fm roughly 
complies with the STAR centrality selection of 20-50\% \cite{STAR:2017ckg}. 
Glauber simulations of Ref. \cite{Abelev:2008ab} were used to relate 
the experimental centrality and the mean impact parameter. 
In the 3FD model, the colliding nuclei have a shape of sharp spheres without 
the Woods-Saxon diffuse edge. This fact, implemented in the Glauber simulations,  
results in a shift of the impact parameter
to a value lower by $\approx$1.5 fm as compared with the results of  
Ref. \cite{Abelev:2008ab}. The nuclear overlap calculator \cite{web-docs.gsi.de} was used 
for this estimate.

%

The polarization in Fig. \ref{fig1} is calculated 
precisely in the same way as 
it is described in Ref. \cite{Ivanov:2020wak}. It is related to the midrapidity 
region $|y_h|\lsim 0.5$, where rapidity range is calculated based on hydrodinamical 
velocities. This constraint roughly simulates the STAR acceptance in terms of 
pseudorapidity $|\eta|<1$. Of course, the acceptance in different experiments may be 
different, the STAR acceptance is implemented just for definiteness. 
The above constraint means that at the freeze-out instant 
the polarization is averaged  
over the central region of colliding nuclei confined by the condition $|y_h|\lsim 0.5$.  
As this condition is related to a certain choice of spatial width of this central 
region, its rapidity width $\Delta y_h$ slightly depends on the collision energy, 
centrality ($b$) and EoS, such that $\Delta y_h/2 \approx 0.5$. 
$\Delta y_h/2$ is not exactly equal to 0.5, e.g., it is 0.45 at $\sqrt{s_{NN}}=$ 2.42 GeV. 
For details of this calculation, please, refer to Ref. \cite{Ivanov:2020wak}. 
The failure of the hadronic EoS at $\sqrt{s_{NN}}>$ 8 GeV is discussed in Ref. 
\cite{Ivanov:2020wak} in detail. To avoid this discussion that is irrelevant to the low energies 
considered here, the hadronic-EoS results are displayed only below the energy of 8 GeV.

The global polarization predicted by the crossover and 1PT EoS's is very 
similar at $\sqrt{s_{NN}}>$ 3 GeV, while at $\sqrt{s_{NN}}<$ 3 GeV 
the predictions differ. 
Predictions of the hadronic EoS slightly differ from two above even at 
3$<\sqrt{s_{NN}}<$ 7 GeV. 
The reason is that the hadronic sectors of the crossover, 1PT and hadronic EoS's 
are similar but not identical. Therefore, the difference between results of different EoS's   
characterizes uncertainty of the present prediction. 
The numerics of simulations with the crossover EoS becomes unstable at $\sqrt{s_{NN}}<$ 3 GeV. 
 Apparenly, this is a consequence of 
somewhat poor quality of tabulation of this EoS at low densities. 
This could be a reason of the large difference of the crossover results from other predictions.
Nevertheless, these low-energy crossover results are displayed in Fig. \ref{fig1} by short-dashed lines.

As seen from Fig. \ref{fig1}, the global polarization increases with the collision energy 
decrease. Depending on the EoS, it reaches values of 6--8\% at $\sqrt{s_{NN}}=$ 3.3 GeV
at $b=$ 8 fm. 
At $\sqrt{s_{NN}}\approx$ 3 GeV this increase slows down or even a maximum is reached, again
depending on the EoS and centrality. This is in agreement with findings in Ref. \cite{Deng:2020ygd}.   
Note the the decrease of the global polarization with the energy rise was predicted in  
Ref. \cite{Rogachevsky:2010ys} long before the first experimental results. That prediction 
was based on the chiral vortical effect.

The appearance of a maximum in the global polarization is very natural. On the one hand, the polarization 
tends to zero when $\sqrt{s_{NN}}$ approaches to $2 m_N$ (two nucleon masses) because the total angular 
momentum of the colliding system becomes close to zero. on the other hand, the polarization rises with 
increase of the collision energy and hence the total angular momentum. For the total polarization, i.e. 
that averaged over the whole range of rapidities, this is a monotonous increase 
\cite{Ivanov:2019ern,Ivanov:2019wzg}. However, the global polarization is measured in the midrapidity 
region. As found both experimentally and theoretically the the global polarization in the BES-RHIC region 
decreases with the energy rise. This happens because the vorticity and hence the polarization migrates to 
backward-forward rapidities 
\cite{Ivanov:2019ern,Ivanov:2019wzg,Ivanov:2018eej,Ivanov:2017dff,Baznat:2015eca,Baznat:2013zx}. 
This migration produces the maximum in the global polarization. The position and the height of 
this maximum depends on the width of the midrapidity window and on the dynamics of the vorticity 
migration to the peripheral rapidities.

Preliminary result by the HADES Collaboration \cite{Kornas:2020qzi}  
was measured at $\sqrt{s_{NN}}=$ 2.4 GeV and the centrality of 10-40\%. 
It best of all corresponds to $b=$ 6 fm among the calculated 
impact parameters. Although this result is still not quite reliably extracted against the background, 
it does not contradict the 3FD prediction at $b=$ 6 fm.

\begin{figure}[bht]
\includegraphics[width=8.5cm]{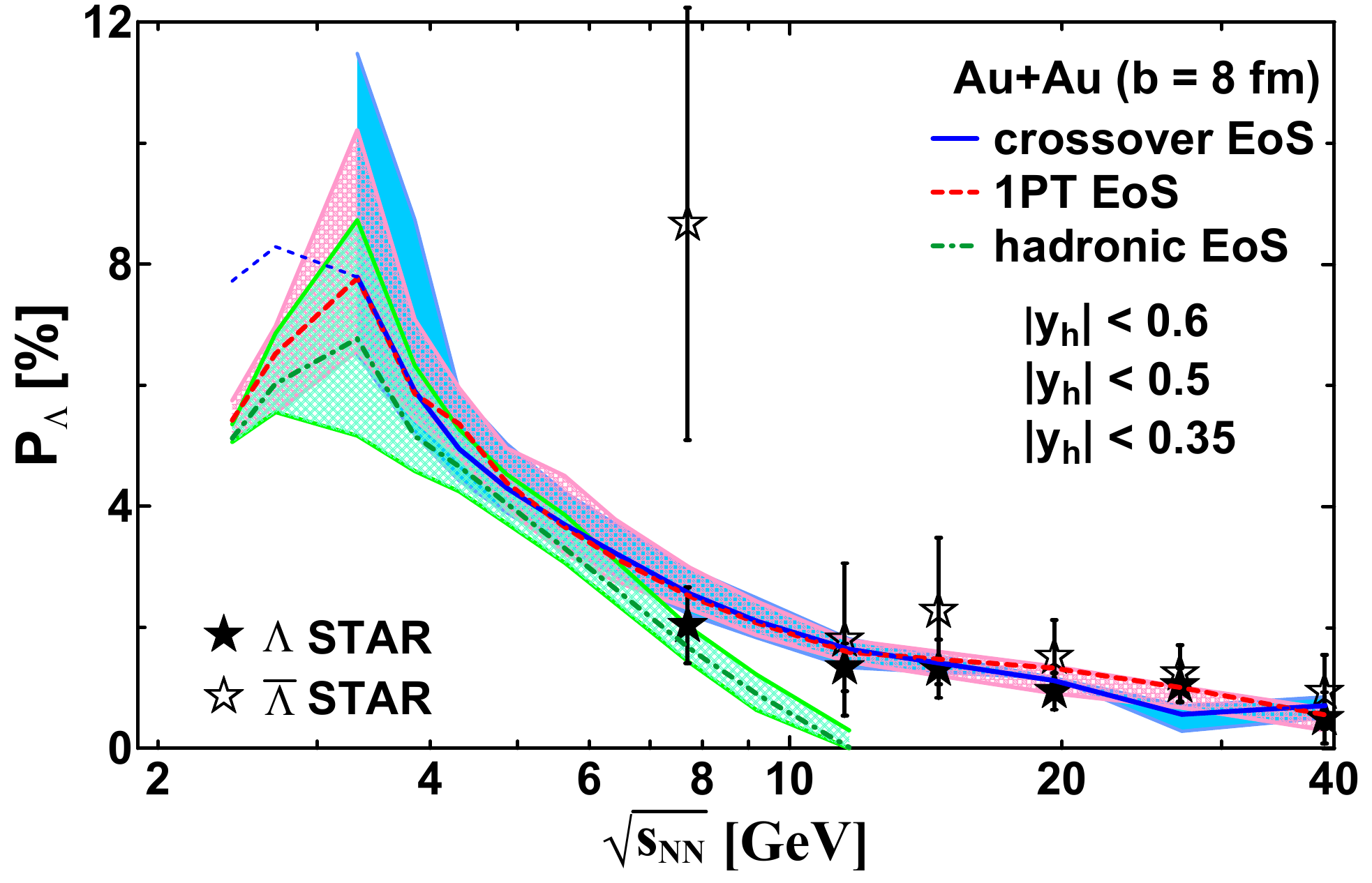}
 \caption{(Color online)
 Global polarization of $\Lambda$ hyperons in Au+Au collisions at $b=$ 8 fm as 
function of collision energy $\sqrt{s_{NN}}$ calculated with different EoS's.    
The lines correspond to the midrapidity region constrained by the $|y_h|\lsim 0.5$ condition. 
The upper borders of the correspondingly colored bands around these lines 
correspond to the constraint $|y_h|\lsim 0.6$, while the lower borders -- to $|y_h|\lsim 0.35$. 
STAR data on global $\Lambda$ and $\bar{\Lambda}$ polarization \cite{STAR:2017ckg} are also displayed. 
}
\label{fig2}
\end{figure}

\begin{figure}[bht]
\includegraphics[width=6.4cm]{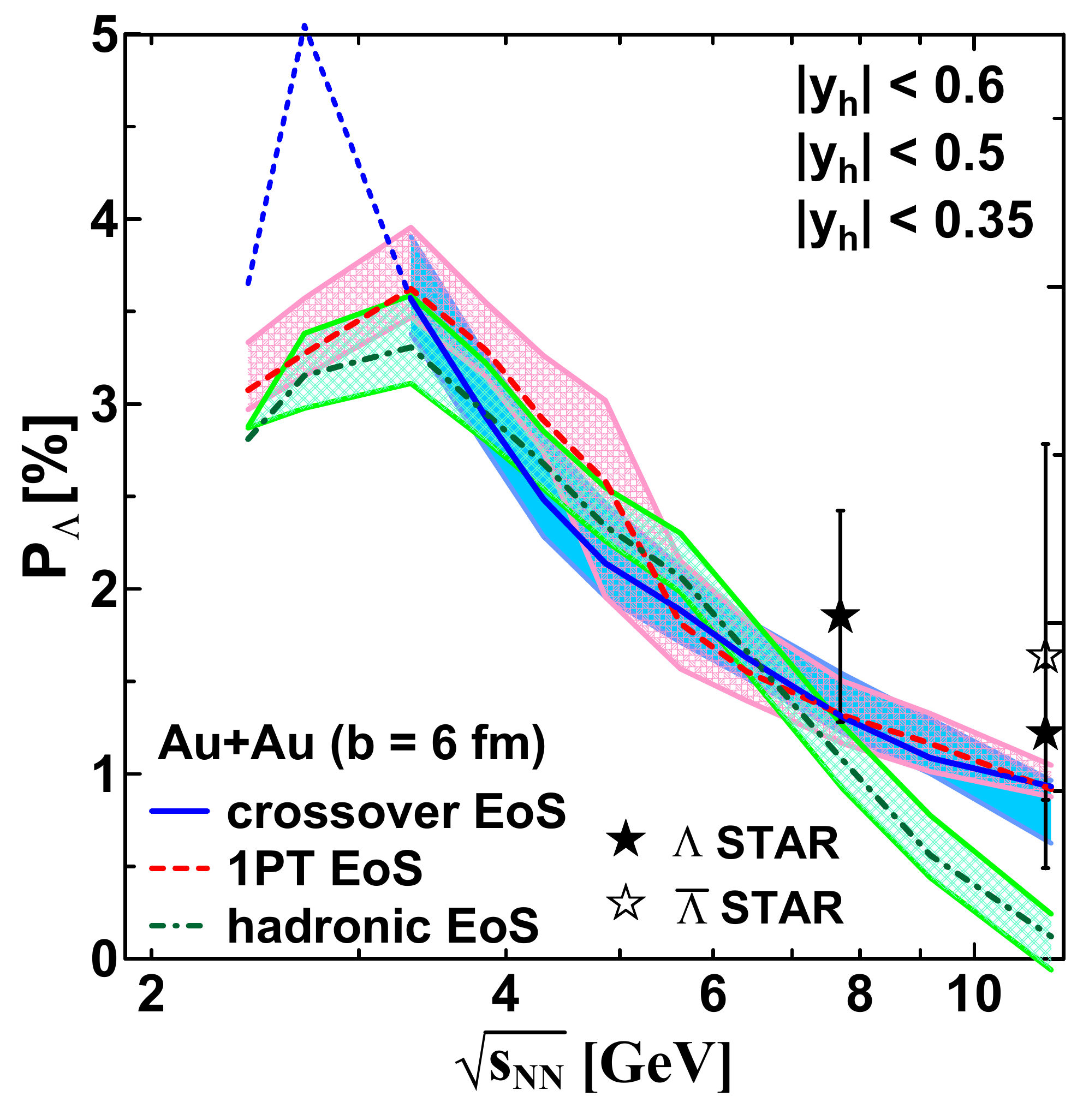}
 \caption{(Color online)
The  same as in Fig. \ref{fig2} but for $b=$ 6 fm.
}
\label{fig3}
\end{figure}

\begin{figure}[bht]
\includegraphics[width=6.5cm]{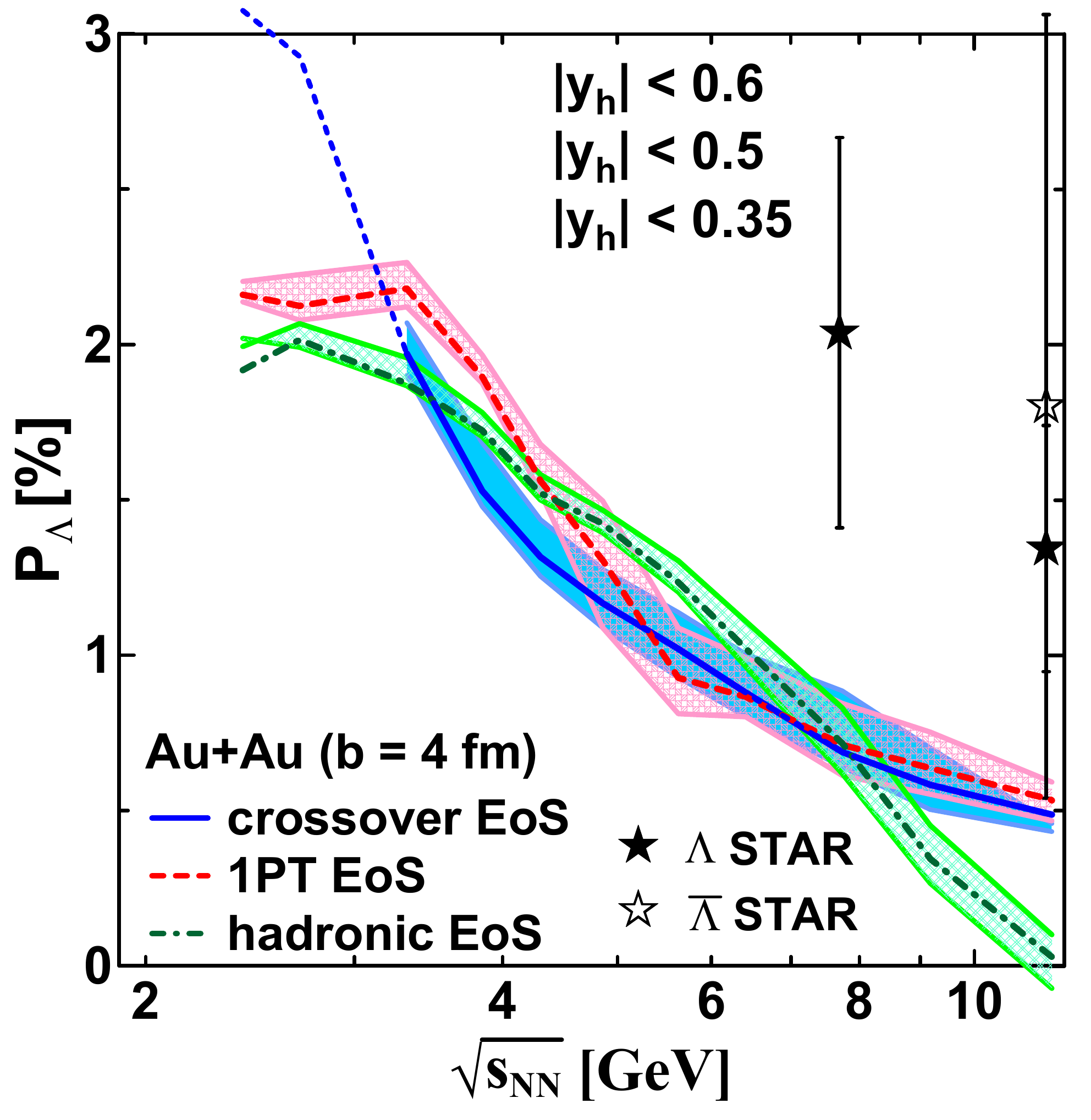}
 \caption{(Color online)
The  same as in Fig. \ref{fig2} but for $b=$ 4 fm.
}
\label{fig4}
\end{figure}

The effect of the rapidity acceptance is demonstrated in Fig. \ref{fig2}, where results at $b=$ 8 fm
are presented for rapidity windows: $|y_h|\lsim 0.5$ (displayed lines), $|y_h|\lsim 0.6$ 
(upper borders of the correspondingly colored bands) and $|y_h|\lsim 0.35$ (lower borders of these bands). 
The global polarization is presented in a wider energy range, 2.4 $\geq\sqrt{s_{NN}}\geq$ 39 GeV, to see 
the effect at various collision energies. At higher collision energies, the effect is small because even 
the wider rapidity range, $|y_h|\lsim 0.6$, amounts a small fraction of the total one. The effect becomes 
stronger with the energy decrease, reaching its maximum at $\approx 3$ GeV. Surprisingly, the effect again 
becomes small at 2.4 GeV. Similar results for other centralities, $b=$ 6 and 4 fm, are presented in 
Figs. \ref{fig3} and \ref{fig4}, respectively. The STAR data are presented there as benchmarks 
for comparing the scale at different centralities. 
The crossover bands are displayed only above 3.3 GeV, because of unstable numerics 
at lower energies.

%
%
%
%
%

In the present calculation the same freeze-out for all species is used. As a result 
the $\bar{\Lambda}$ polarization turned out to be 
very close to the $\Lambda$ one, therefore it is not presented here.
Earlier freeze-out of $\bar{\Lambda}$'s partially solves the problem of the 
$\Lambda$--$\bar{\Lambda}$ splitting \cite{Vitiuk:2019rfv}, but not at the 
lowest BES-RHIC energy of 7.7 GeV. 

Feed-down contribution to the $\Lambda$ polarization due to 
decays of higher mass hyperons is not taken into account in the present estimate. 
This feed-down can reduce the polarization by about 10--15\%, as demonstrated in Refs. 
\cite{Karpenko:2016jyx,Becattini:2020ngo,Becattini:2016gvu,Becattini:2019ntv,Xia:2019fjf}, 
albeit at higher collision energies. 
This feed-down effect is definitely of minor importance at low collision energies 
because of low abundances of higher mass hyperons.

An alternative approach to the global $\Lambda$ polarization 
is based on the Axial Vortical Effect
(AVE) \cite{Vilenkin:1980zv,Son:2004tq,Gao:2012ix,Sorin:2016smp},
which is in fact identical to the Chiral Vortical Effect (CVE).   
It reasonably good describes the data on the global polarization \cite{Baznat:2017jfj,Ivanov:2020qqe}
and naturally explains  the above mentioned $\Lambda$-$\bar{\Lambda}$ splitting
\cite{Baznat:2017jfj,Ivanov:2020qqe}.
However, in the energy range below the NICA region 
it is hardly applicable because the chiral symmetry, i.e. the driving force of the AVE, 
is spontaneously broken.

To summarize, based on the 3FD model, predictions for the global $\Lambda$ polarization 
in Au+Au collisions in current and upcoming experiments at moderately relativistic energies, 
2.4 $\leq\sqrt{s_{NN}}\leq$ 11 GeV, are made. It is predicted that the polarization reaches a maximum 
or a plateau (depending on the EoS and centrality) at $\sqrt{s_{NN}}\approx$ 3 GeV. 
It is found that the global polarization increases 
when the width of the rapidity window around the midrapidity rises.
The preliminary result of HADES Collaboration \cite{Kornas:2020qzi}, 
although not yet quite reliably extracted against the background, 
does not contradict the results of the present calculation.
Of course, not all related aspects were considered in this short paper, e.g. 
a more detailed study of the rapidity dependence of the global polarization is required, 
applicability of the thermodynamic approach to 
nuclear collisions at low energies (the collision dynamics 
becomes less equilibrium with the collision energy decrease) should be checked, etc. 
These topics will be considered elsewhere.

Helpful discussions with 
F. Becattini, O.V. Teryaev, V.D. Toneev,  
and D.N. Voskresensky are gratefully acknowledged. 
This work was carried out using computing resources of the federal collective usage center ``Complex for simulation and data processing for mega-science facilities'' at NRC "Kurchatov Institute", http://ckp.nrcki.ru/.
This work was partially supported by the Russian Foundation for
Basic Research, Grants No. 18-02-40084 and No. 18-02-40085, and  
by  the Ministry of Education and Science of the Russian Federation within  the Academic Excellence Project of 
the NRNU MEPhI under contract 
No. 02.A03.21.0005.  

\end{document}